\documentclass[a4paper,12pt]{article}
\usepackage{graphicx} 
\usepackage{etoolbox}

\makeatletter
\patchcmd{\abstract}{\quotation}{\quotation\small}{}{}
\makeatother

\usepackage[margin=1in]{geometry}
\usepackage[backend=biber,style=authoryear]{biblatex}
\usepackage[labelfont=bf]{caption}
\usepackage[nobottomtitles]{titlesec}
\usepackage{footmisc}
\usepackage{needspace}
\usepackage{float}
\usepackage[hidelinks, colorlinks=true, linkcolor=blue, urlcolor=blue, citecolor=blue]{hyperref}
\title{EJRA_Critique}
\date{June 12th, 2024}

\begin{document}

\begin{titlepage}
	\thispagestyle{empty} 
	\title{Is the EJRA proportionate and therefore justified? A critical review of the EJRA policy at Cambridge}
	\author{Oliver Linton, Raghavendra Rau, Patrick Baert, Peter Bossaerts,\\
		Jon Crowcroft, G.R. Evans, Paul Ewart, Nick Gay,\\
		\newline  Paul Kattuman, Stefan Scholtes, Hamid Sabourian, and Richard J. Smith \\
		Universities of Cambridge and Oxford}
	\maketitle
	\begin{abstract}
		This paper critically evaluates the HESA (Higher Education Statistics Agency) Data Report for the Employer Justified Retirement Age (EJRA) Review Group at the University of Cambridge (\cite{CambridgeHESA2024}), identifying significant methodological flaws and misinterpretations. Our analysis reveals issues such as unclear application of data filters, inconsistent variable treatment, and erroneous statistical conclusions. The Report suggests that the EJRA increased job creation rates at Cambridge, but we show Cambridge consistently had lower job creation rates for Established Academic Careers compared to other Russell Group universities, both before and after EJRA implementation in 2011, with no evidence for a significant change in this deficit post implementation. This suggests that EJRA is not a significant factor driving job creation rates. Since other universities without an EJRA exhibit higher job creation rates, this suggests job creation can be sustained without such a policy. We conclude that the EJRA did not achieve its intended goal of increasing opportunities for young academics and may have exacerbated existing disparities compared to other leading universities. We recommend EJRA be abolished at Cambridge since it does not meet its aims and could be viewed as unlawful age discrimination.
		\newline
		\it{This version is a revision reflecting some of the comments made by members of the university EJRA review group in public discussion, see \cite{Holmes24}.}  
	\end{abstract}

	{\small \noindent\emph{Keywords}: Discrimination, Job Creation; Retirement; Treatment Effect, Universities\par}
	{\small \noindent\emph{JEL Classification}: C12, C21, C23, J21, J26, J63, J83.\par}
\end{titlepage}

\setcounter{page}{1}

\section{Introduction}

The University of Cambridge employs an Employer Justified Retirement Age
(EJRA) policy, mandating retirement for all academic and academic-related
staff at the end of the academic year in which they turn 67. Following the
abolition of the Default Retirement Age (DRA) by the UK Coalition government
in 2011 following on from the Equality Act (2010), employers were permitted
to impose mandatory retirement only on the condition that it could be
objectively justified as a proportional means of achieving a legitimate
social policy aim. Implemented in 2012, after consultation with its staff,
the Cambridge EJRA has four stated goals:

\begin{itemize}
\item To ensure inter-generational fairness and career progression; 

\item To enable effective succession-planning; 

\item To promote innovation in research and knowledge creation; and 

\item To preserve academic autonomy and freedom.
\end{itemize}

While the twenty two other Russell Group universities in the U.K. did not
choose to introduce an EJRA in 2011, Cambridge and Oxford universities did
so, setting them apart from the other Russell Group universities (and indeed
other employers).

The effectiveness of the EJRA in achieving its aims is predicated upon its
effect upon the rate at which vacancies arise from retirement and other
causes. Estimates of the effect of the EJRA on the vacancy creation rate
(VCR) may be made using models employing simplifying assumptions. A simple
model, presented by Paul Ewart during debates in Congregation at Oxford%
\footnote{%
See \url{https://tinyurl.com/yc68vksc}, page 443.}, assumes a fixed number
of posts subject to the EJRA and a uniform age distribution i.e. equal
numbers of staff in each year group by age.\footnote{%
The Data Report, \cite{CambridgeHESA2024}, confirms that this reflects the
age distribution also at Cambridge.} Then, initially assuming that vacancies
arise \textit{only} by retirements at the EJRA when a cohort reaches the
retirement age and are filled by younger academics at the average age of
appointment, the process can be viewed as a queue where the numbers joining
equal the numbers leaving by retirement. If the EJRA was abolished, then the
average age at which people retired would increase, at most, by about 3
years, according to survey data. In the steady state, the length of the
queue would increase, but the number in each age group would decrease
correspondingly to maintain the same number of staff. Hence the number of
vacancies each year would decrease by the same amount.

The effect of an EJRA then is to increase the Vacancy Creation rate (VCR)
relative to the situation with no mandatory retirement. It is on this basis
that the University seeks to justify its EJRA policy as a means to achieving
its stated aims. However, since the policy discriminates on the basis of a
protected characteristic, age, it is unlawful unless it can be objectively
justified as a \textbf{proportionate} means of achieving those aims. Using
this simple queuing model, it is possible to
estimate the effect of the EJRA on the VCR as a measure of its
proportionality.

This article critically evaluates the Data Report, \cite{CambridgeHESA2024},
identifying significant methodological flaws and misinterpretations. The
Data Report suggests that EJRA at the University of Cambridge increased job
creation rates. Our analysis reveals issues with inappropriate data filters,
inconsistent variable treatment, and erroneous statistical conclusions. In
fact, Cambridge consistently had lower job (vacancy) creation rates for
Established Academic Careers (EAC) compared to other Russell Group
universities, both before and after the EJRA implementation. Furthermore,
they supply no credible evidence for a significant change in this deficit
after implementation. This implies that EJRA is \textit{not} a significant
factor driving job creation rates. Since other Russell Group universities,
which do not have an EJRA, exhibit higher job creation rates, this suggests
that job creation can be sustained or even thrive without such a policy.

\section{Qualitative assessment of the EJRA}

Using the data quoted in the Data Report, the average age at appointment of
40 and retirement at the EJRA at 67 gives an average career length of 27
years. Assuming, on the basis of survey and other data, that on average
people will work for a further 3 years, the average career length will
increase by 3/27 i.e. 1/9 or about 11\% and the VCR will decrease by a
corresponding proportion. Therefore, for a fixed number of posts, the EJRA
increases the VCR by about 11\%. However, as confirmed by the Data Report,
at least 50\% of vacancies arise for reasons other than retirement, so the
effect of the EJRA on the VCR is reduced to 5.5\%. In addition, about 50\%
of those reaching the EJRA will retire voluntarily and so the relative
effect of the EJRA on the VCR is reduced still further to about 2.75\%.

The Ewart Employment Tribunal judged that an increase in vacancy creation
rate of between 2-4\% was trivial
and not a proportionate means of achieving the aims.\footnote{%
See \url{https://tinyurl.com/2ywaft57}} This finding was upheld by the
Employment Appeal Tribunal\footnote{%
See \url{https://tinyurl.com/3eea3bkm}} and was further confirmed in a
recent ET cases (2023) involving four claimants at Oxford University.%
\footnote{%
See \url{https://tinyurl.com/447ecv5a}, \url{https://tinyurl.com/5ahknzkk},  %
\url{https://tinyurl.com/5efvjakv}, and \url{https://tinyurl.com/3zxt2kd7}%
\label{Ewartcases}} It was acknowledged by Oxford University's statistical analysis in these recent ET tribunals that the
modelling method presented by Ewart "gave similar results to the
methodology based on Little's Law of queuing, once the
correct assumptions were used".\footnote{%
Witness statement of Malgorzata Turner in Employment Tribunals referred to
in footnote \footref{Ewartcases} above.}

It is worth noting that the recommendation of the Review Report that the
EJRA should be changed to set at 69 instead of 67 will result in the policy
being even less proportionate. This is because the average extension of 3
years will be to a career of 29 years, rather than 27 years, a percentage
reduction of the VCR due to retirements alone from 11\% (3/27) to 10\%
(3/29) and an overall reduction in the VCR of 2.5\%. However, if people
continue to retire at the current average age, then the extension will be
only 1 year past 69, giving a career extension of 1/29 or 3\% and an overall
reduction of the VCR of 0.75\% when vacancies for other reasons and
voluntary retirements at the EJRA are taken into account.

The "steady state" conditions of this simple model
do not take account of the transient effect induced by a sudden abolition of
the EJRA or a change in the age from say 67 to 69. In the latter case, and
using the same assumption of a uniform age distribution, there would be no
retirement-generated vacancies for two years after which the situation would
evolve into the steady state over time. This effect is treated in a dynamic
modelling approach using Little's Law that would also have to take account
of the vacancies created by reasons other than retirement and the average
duration of extensions of employment post-EJRA.

It is clear, however, given the similarity in the situations at Oxford and
Cambridge; same age at appointment of 40 and 50\% of vacancies arising from
reasons other than retirement etc., one expects that the increase in the
vacancy creation rate by the EJRA at Cambridge will also be in the range 2 
-4\%. Any significant deviation from this
estimate would suggest that some filtering of the statistics data has been
made to change the assumptions built into the model.

We next turn to the specific goals of the EJRA and examine whether the EJRA
is a proportionate means of achieving those goals. Recall that the EJRA sets
four goals: \newline

\textbf{Promoting Innovation and Knowledge Creation}

The EJRA is predicated on an antiquated university model with a fixed number
of professorships, necessitating one professor's departure for another's
ascension. This premise aligns with the "lump of labor fallacy",  the erroneous belief that there is a fixed amount of work to be
distributed, leading to the misconception that increased productivity causes
unemployment. However, this notion is fallacious (\cite{Krugman2024}).
Arguing against productivity enhancements due to potential job losses is
irrational. The creation of new professorships is contingent on funding
availability and academic interest, not vacancy generation through
retirements. Hiring new faculty members can potentially stimulate novel research avenues and ways of thinking, fostering innovation and knowledge creation, rather than merely replacing outgoing scholars. It is ageist prejudice, however, that assumes older academics do not make new or important contributions. Research has shown, for example, that a scientist's most impactful contribution can come at any stage of his or her career, \cite{Sinatra2016}. In addition, the senior outgoing scholars are likely able to generate significant amounts  of funds, all of which are lost to the university. \newline

\textbf{Enabling Effective Succession Planning} It is acknowledged that at
least 50\% of vacancies arise before the mandatory retirement age and these
are managed successfully, even if they do not occur at predictable times. It
is therefore difficult to imagine that vacancies arising from voluntary
retirement could not be equally well managed. Indeed, it is likely that
academics wishing to retire would be willing and capable of giving adequate
prior notice of their intention to retire and so mitigate any potential
issues related to succession planning.

The EJRA hinders the University's ability to attract senior external
candidates, as the prospect of mandatory retirement within a few years can
deter potential applicants. Moreover, the policy impedes academic
productivity well before retirement, as academics may be unable to supervise
doctoral students or apply for grants at least five years prior to the
retirement age due to the impending termination of their employment.
Consequently, productive senior academics often seek employment elsewhere
well before reaching the EJRA, resulting in a detrimental brain drain of
highly sought-after and accomplished scholars from the University. This has
the impact of reducing predictability since senior academics seeking
positions elsewhere will be unlikely to notify the university ahead of time
that they are actively looking elsewhere. \newline

\textbf{Ensuring Intergenerational Fairness and Career Progression}

It is improbable that junior academics at Cambridge directly benefit from
the forced retirement of their senior colleagues. Senior positions are
advertised globally, and the potential for internal promotions is often
outweighed by external applicants. Thus, even if the model underpinning the
Review Report is sound (which it clearly isn't), the intergenerational
fairness argument would not apply within the University context but rather
on a global scale, where Cambridge's retirees ostensibly create
opportunities for younger academics worldwide. Needless to say, this would
only be a drop in the ocean and by no means a proportionate measure.
Curiously, this rationale does not extend to administrative positions, which
are exempt from the EJRA. Additionally, the policy disproportionately
disadvantages women, whose career trajectories are often prolonged due to
family obligations, resulting in a shorter time frame before reaching the
retirement age after achieving career milestones. In any case, the increase in the number of vacancies each year attributable to the EJRA is trivial - of the order of a few percent 
\begin{figure}[tbp]
\centering
\includegraphics[width=0.7\textwidth]{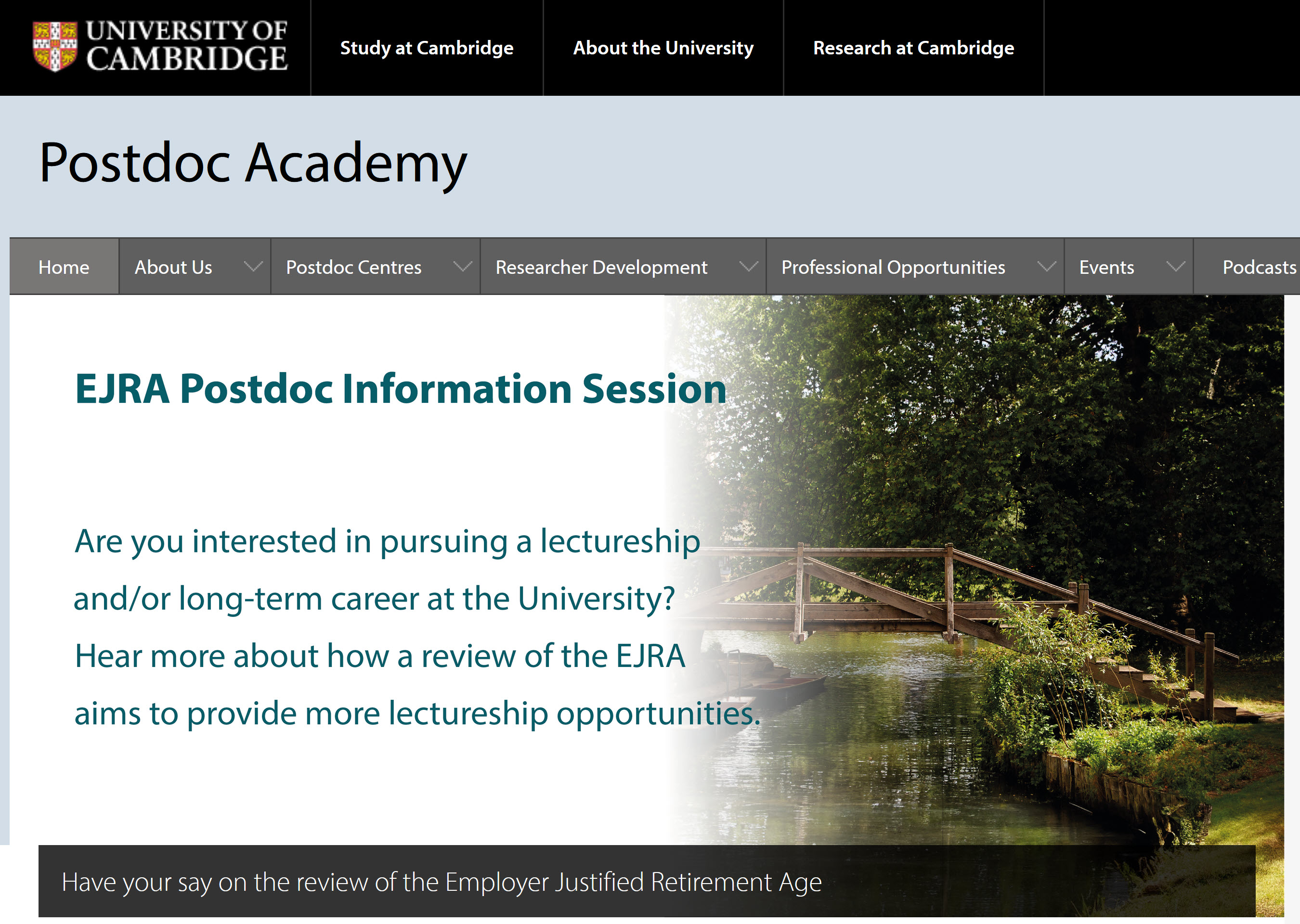}  
\caption{University website for postdoc students at Cambridge: Retrieved 20
May 2024}
\label{fig:CamPostDoc}
\end{figure}

\begin{figure}[]
\centering
\includegraphics[width=0.9\textwidth]{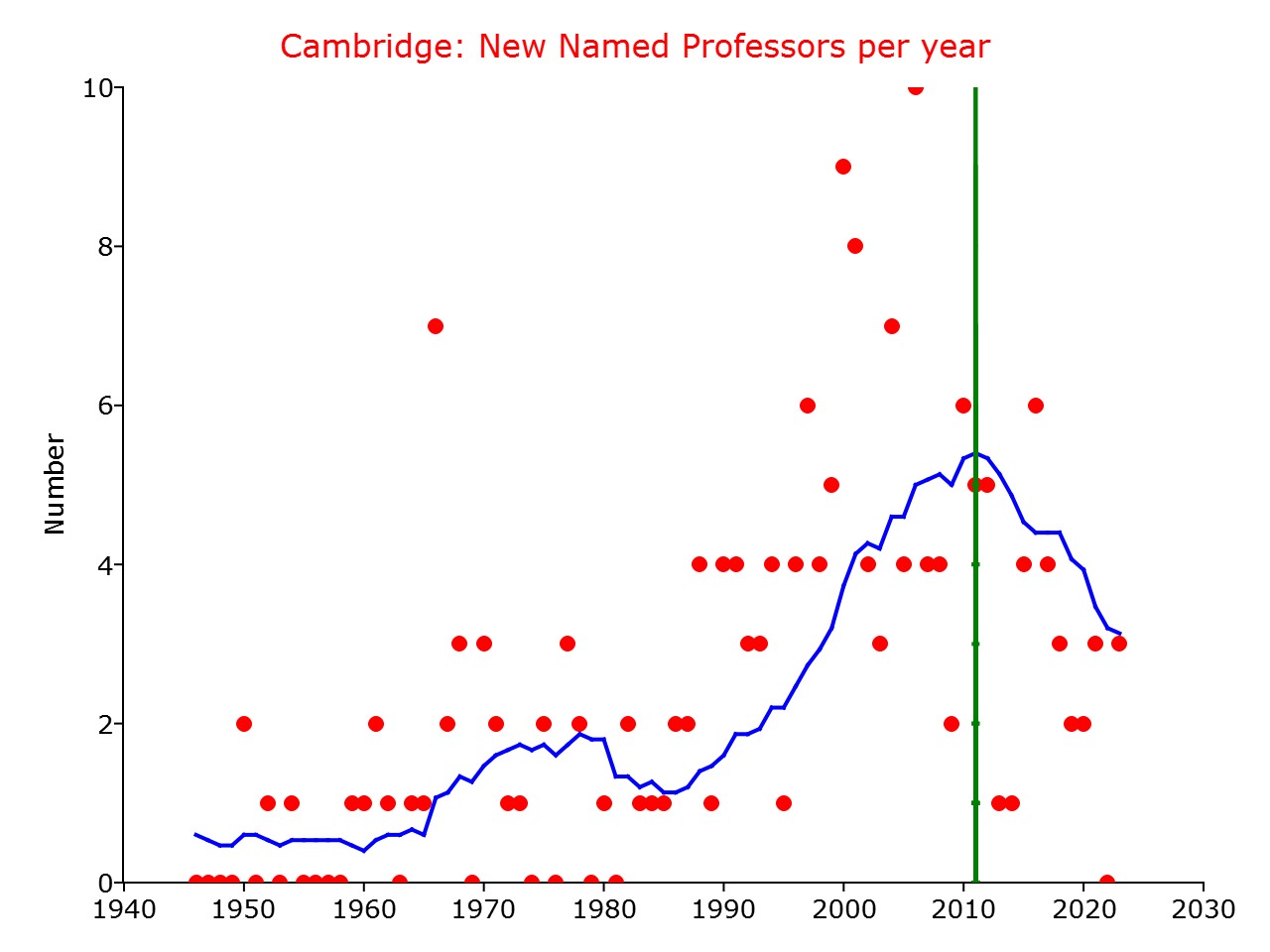}  
\caption{Number of professorships created at Cambridge (15 year trailing
average)}
\label{fig:CamProfs}
\end{figure}

In Figure \ref{fig:CamPostDoc}, the university argues that the EJRA policy
promoted in the HESA review \textit{increases} the lectureship opportunities
available to post-doctoral colleagues (\cite{PostDocWebsite}). Since the
Data Report suggests \textit{increasing} the retirement age by two years to
69, it is difficult to understand how lectureship opportunities will be
increased. One possibility is that the University plans on increasing the
number of senior positions such as professorships. Unfortunately, a simple
graph of the number of senior professorships at Cambridge, depicted in
Figure \ref{fig:CamProfs} shows a very sharp decline in these positions
after the EJRA period in Cambridge.

\textbf{Preserving Academic Autonomy and Freedom} Finally, the EJRA also
undermines academic autonomy and freedom. To secure an extension beyond the
retirement age, academics must demonstrate productivity, which may
incentivise risk-averse behavior and incremental research rather than
pursuing potentially groundbreaking but uncertain avenues. This dynamic has
the potential of stifling innovation and knowledge creation, which should be
a core objective of a leading research university. \newline

In summary, these arguments undermine the claims in the Review Group’s Report, that the EJRA contributes significantly to the aims of intergenerational fairness, innovation, succession planning or academic freedom and, on the contrary, may have adverse effects upon its pursuit of excellence in research and education. 

\section{Quantitative assessment of the Data Report}

The Data Report sought to establish that EJRA has a beneficial effect on
vacancy creation based on historical data and statistical techniques. In
general terms and when properly applied, the method of analysis
(difference-in-differences) used is capable of measuring the impact of a
policy by comparing changes, from before to after the policy change, between
a group that experienced the policy and a group that did not. In the
analysis of whether the removal of EJRA at Cambridge would affect job
creation, one group was the RRG (Remaining Russell Group) where EJRA was
removed, and the other group consisted solely of Cambridge where an EJRA was
introduced so that mandatory retirement continued, and the
difference-in-differences calculation was applied to vacancy creation rates
in the RRG and in Cambridge, before and after EJRA removal in the RRG. This
section argues that Cambridge's HR data was apparently used to design the
filter that was then applied to the HESA data of all universities, and yet
full details of this method were not provided, which is contrary to sound
practice. The analysis as applied did not adequately control for
pre-existing trends, or for differences between universities. In fact,
Cambridge consistently had lower vacancy creation rates compared to other
universities, both before and after EJRA removal in the RRG. It is shown
below that the conclusion that the removal of the EJRA policy would
significantly impact the vacancy creation rate in Cambridge simply does not
follow from the analysis.

The second part of the Data Report uses a system dynamics model to simulate
forward what would be the likely consequences of the abolition of the EJRA
at Cambridge in terms of vacancies. In the second part of this section, we
point out significant limitations in that analysis as well.

\subsection{Effect of EJRA on Job Creation Rate}

The authors of the Data Report obtained data from HESA on the number of
employees in different categories for each year from 2007-2008 academic year
to 2021-2022 academic year for each of the 24 Russell Group universities
including Cambridge and Oxford. Oxford is excluded from the analysis for
political/legal rather than scientific reasons with a comparison made only
between Cambridge and RRG. The authors describe on page 7 how they applied
various ``filters" to the HESA data using Cambridge HR data to construct the
group that is eligible for the EJRA (which was not identified from the HESA
data alone). Since they do not provide details, it is unclear how they chose
the filter to match the Cambridge HESA data with the Cambridge HR data. They
then applied that same filter to all of the HESA data for the RRG group as
well. They did not and could not obtain HR data from the RRG universities
and so again, it is not completely clear that a fair comparison was made
given the differences across these diverse universities.  \cite{Holmes24} reports that "\textit{ the filtered dataset covers 93\% of the relevant staff cohorts and excludes all staff not subject to EJRA}". In fact, before 2011 everyone in Cambridge and RRG were subject to the national Default Retirement Age and so it is not so clear why attention is restricted to such a subset of the work force, and whether the exclusionary criteria have the same meaning at different universities. 
For example, salary point 38 and above is likely to mean something quite different at the LSE
versus Exeter University and versus Cambridge. It also matters whether this so-called filter was preregistered or whether it was introduced because the authors' calculations without the filter were not so impressive. 

The Data Report examines two classes of staff at Cambridge: Established
Academic staff (EAC) and Established Academic-related staff (EAR). The EAC
group includes academic staff whose primary function is teaching and/or
research. Specifically, the Data Report defines EAC as employees registered in the HESA dataset with:

\begin{itemize}
\item Academic employment function coded as 1 (Teaching Only), 3 (Teaching 
and Research), or 9 (Neither, e.g. Vice-Chancellor)

\item Salary point 38 and above

\item Full-time or full-time term-time only employment mode

\item Open-ended/permanent terms of employment
\end{itemize}

The EAR group includes non-academic professional staff supporting teaching
and research. The Data Report defines EAR as employees with:

\begin{itemize}
\item Academic employment function coded as 4 (Not an academic contract) or 
X (for data before 2011/12)

\item Salary levels point 33 and above

\item Wholly or partly financed by the higher education provider

\item Open-ended/permanent terms of employment
\end{itemize}

In effect, the EAC group represents the core group of established academic
teaching and research staff, while the EAR group covers the non-academic
professional support staff, both at relatively senior levels within the
university. The purpose of the Data Report is to show that the university
created a significantly higher number of jobs in the Cambridge EAC group
than would have arisen if Cambridge had not had an EJRA by comparing with
other universities that did not operate an EJRA over the relevant period.

The main variable of interest is defined as 
\begin{equation}
r_{it}^{jc}=\frac{N_{it}^{jc}}{N_{it}},
\end{equation}%
where $N_{it}$ is the total employees in a given group (for example,
university $i)$ at year (actually academic year) $t,$ while $N_{it}^{jc}$ is
the total number of new appointees at university $i$ in year $t$ (presumably
new appointees in the specific group). This is the measure of job creation
that is identified as the outcome of the EJRA policy.

\begin{figure}[tbp]
\centering
\includegraphics[width=1\textwidth]{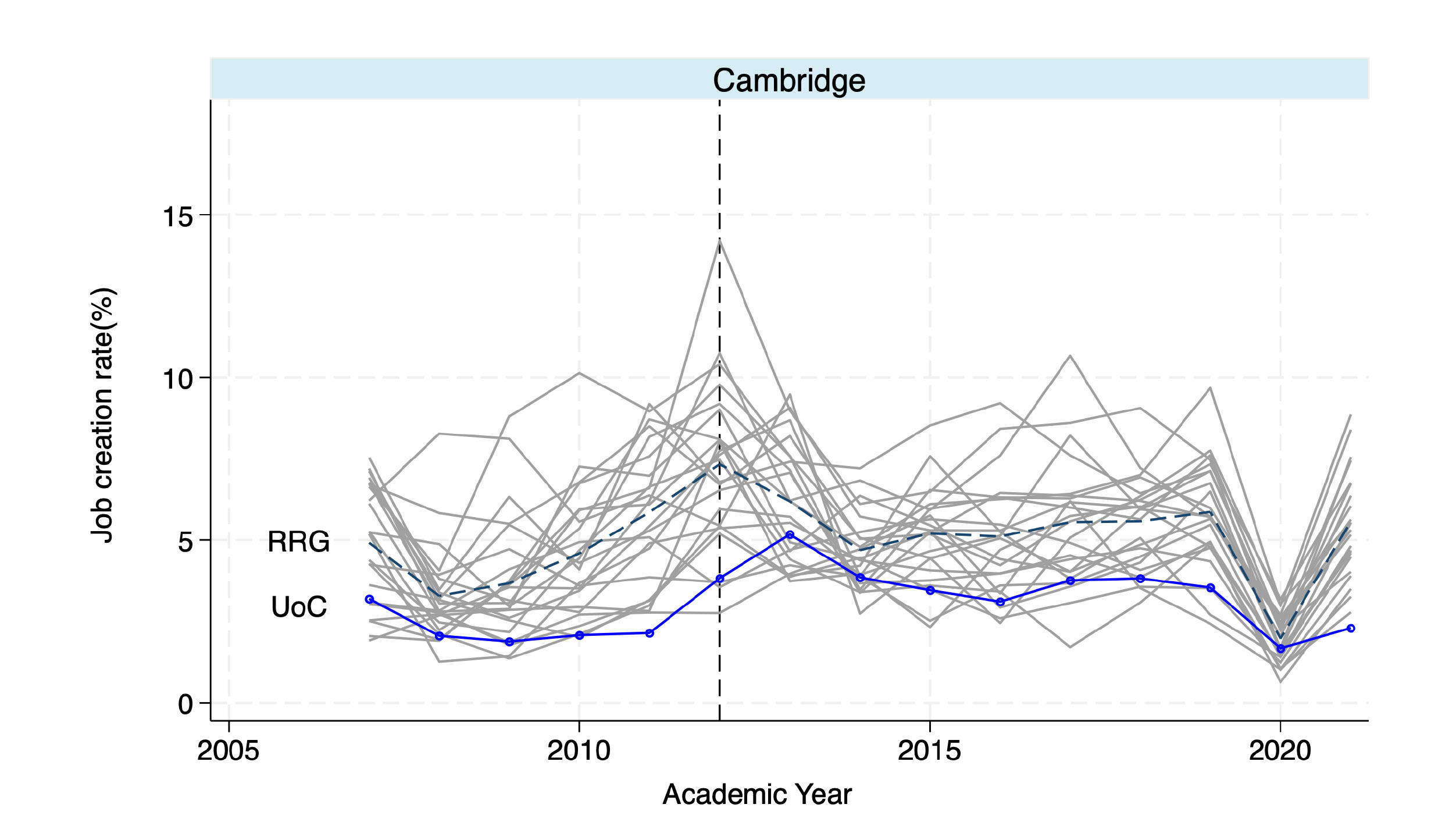}  
\caption{Figure A1 reproduced from the Data Report (\protect\cite%
{CambridgeHESA2024})}
\label{fig:PentyFigureA1}
\end{figure}

This variable is shown in Figure A1 for both the EAC and EAR groups (see
Figure \ref{fig:PentyFigureA1} reproduced below). 
Figure 3 further breaks this down to EAC STEM and EAC non STEM groups. These
figures show the time series of the 22 RRG universities and Cambridge over
the period 2007-2008 to 2021-2022 (a period of 15 years) along with the
average across RRG and a vertical line indicating 2012, the date from which
the abolition of retirement age at RRG is binding. Regardless of the figure
examined, the key feature is that Cambridge has had a \textit{lower} job
creation rate in comparison with \textit{every} single other RRG university.
This is acknowledged on page 8 of the Data Report, \textquotedblleft \textit{%
UoC has a lower job creation rate for EAC compared to most other HEPs}
(Higher Education Providers). \textquotedblleft \textit{Figure A1 shows the
average job creation rate is 3\% for UoC between AY 2007/08 and 2021/22,
while the average for RRG is 5\%.}" In one sense, the reader can stop here
since the raw data clearly indicates a result completely divergent from its
core message that the EJRA has created significantly more jobs at Cambridge.
However, there are potential flaws in this interpretation that suggest we
look further.

Specifically, in the Data Report they argue that these numbers are the
average over all fifteen years and it is plausible that Cambridge has
improved its performance relatively since 2011, post the introduction of
EJRA. A simple way of capturing this pattern would be to compare the average
(over time) of Cambridge performance after 2011 with the average of the RRG
universities. Unfortunately, looking at Figure \ref{fig:PentyFigureA1}, it
is clear that Cambridge would lose that contest too: In every single year,
Cambridge created fewer EAC jobs than the average of the RRG universities
and in every year, it has been in the bottom 5.

Yet another possibility is that the EJRA improved Cambridge's job creation relative to what would have happened if it had likewise eliminated mandatory retirement. In that case, what we need to do is to
compare the average performance of Cambridge post 2012 with the average
performance of Cambridge pre-2012 and do the same for the RRG universities.
This is the heart of why their argument appears to work since one can see
that just prior to 2012 the RRG universities created a lot of new positions
and their performance subsequent to 2012 was relatively worse. Actually, the
key feature of the pre-2012 data was the big dip following the 2008 Global
Financial Crisis (GFC) and the strong rebound, especially for RRG
universities, just running up to 2012, and for Cambridge starting slower but continuing to 2013. 

The main approach they use is a so-called natural experiment, which is
intended to mimic how clinical trials are carried out. The treatment or
policy variable is the introduction of the EJRA by Cambridge in 2011, and
the treatment group is Cambridge with pre-treatment period
2007/2008-2011/2012 and post treatment period 2012/2013 (it takes a year for
the abolition of retirement to take effect) to 2021/2022. The control group
is the RRG universities. They provide a formal statistical test of whether
this treatment effect is zero (there is no effect) versus the alternative
that there is an effect. It is important to note that this design addresses
the issue about abolition of EJRA as it compares universities without EJRA
with Cambridge that does. It does \textit{not} provide any evidence about
the effect of raising the age of EJRA to 69 from 67. As outlined in the
introduction above, increasing the age of the EJRA results in its being even
less proportionate as a means of achieving the aims.

The main issues for the application of this method here are:

\begin{enumerate}
\item Treatment is not randomly assigned as in a clinical trial. Cambridge,
the authors themselves argue is \textit{different}, (and the authors of the Oxford EJRA argued
the same about Oxford before them) and those differences lead them to
choose the EJRA policy in the first place. Fundamentally, one cannot
separate out whether the difference between Cambridge and the RRG
universities or the change in difference is due to the EJRA or due to the fundamental differences
already present that manifest themselves in: the level of the vacancy rate,
its variability, and how it responds to external shocks.

\item The treatment is really 2012 itself, not the introduction of the EJRA 
(or rather the division is into the pre- and post-2012 periods). The EJRA 
was not the only thing that changed after 2012 compared to before 2012 and 
that could influence vacancy rates differentially across universities.

\item The sample sizes are too small. There is only one treated patient, 
Cambridge, and only 22 untreated patients, the RRG (or equivalently, 22 treated patients and one untreated). Imagine testing a 
vaccine on one patient and applying a placebo to twenty two. Pfizer's 
Covid-19 vaccine clinical trial had over 46,000 participants with thousands in both groups. The so-called statistics
here all rely on large sample approximations, and here the smallest group 
average is made from 5 time series observations.

\item The outcome variable is a ratio that is typically a small positive
number. One gets the impression from reading statistical textbooks (such as 
\cite{CoxHink74}) that such a variable should not be approached by linear
methods.

\item The unit of measurement is at the university level. In a clinical
trial, one is looking at individuals, and it is reasonable to assume that
although they are different in fundamental ways, the health outcomes are
essentially independent across individuals and only depend on the
treatment. It is hard to believe that is the case for a university's
economic decisions, which is what is needed here. For example, all the
universities are admitting undergraduates through UCAS and they are all
trying to recruit from the same pool of academics. The consequence of this
is that there probably is not much statistical gain from looking at the RRG
universities separately rather than just focussing on their average performance, i.e., there is really just one patient and one control.
\end{enumerate}

In conclusion, the methods adopted are not appropriate for the data they
have. The results they obtained are not credible. The fundamental fact is
that Cambridge creates fewer new academic jobs in comparison with RRG
whether or not they both had a retirement age. This is clear from Figure A1.
There is no convincing evidence that in the counterfactual world where
Cambridge had no EJRA that it would generate materially fewer vacancies.

Figure A30 from the Data Report (reproduced in Figure \ref%
{fig:PentyFigureA30}) gives their estimates of the treatment effect
separately for each year along with confidence bands. This is defined as the
difference in job creation between Cambridge and RRG relative to the same
difference for 2012 (the zero value for 2012 is suppressed from the figure).
According to this graph, the introduction of EJRA had a significant positive
relative effect even five years before it was introduced! This can't be
right. In the language of natural experiments, this strongly suggests that
the pre treatment parallel trends assumption is violated, and so their main
results are not valid. This they acknowledge, which leads them to implement
a further couple of methods for dealing with this issue. Specifically, \cite{Holmes24} says "\textit{we prefer the student number adjusted approach}" whose results are reported in Table A4, Column 3. Interestingly, these results show a significant positive effect not just for the EAC group but also for the EAR group, so it is surprising that the main report recommends abolishing EJRA for the EAC group, if this is their preferred result. Unfortunately,
their cure for non parallel trends is not done correctly and raises further issues. In the Appendix we give a more formal treatment of the statistical
issues. 
Another interesting aspect of Figure \ref%
{fig:PentyFigureA30}) is that there is a
declining trend in job creation rates in the post-Brexit period (from 2017
onwards) interrupted by a spike upwards during the Covid year of 2020. Even
taking their methods at face value, by 2021, the effect has totally
disappeared, so whatever was driving their results has vanished. Another
interpretation of this data is just that, as in any time series, there are
periods where the series is above trend and periods where it is below trend
and these cycles are determined by a complex set of reasons that have not
been investigated in their report. The authors of the report argue that EJRA
is the most likely explanation for the positive treatment effect that they
found and that our work does not suggest an alternative explanation. Here,
we mention a few possible alternative explanations for the variation
of these time series over the fifteen years they examine. 

\begin{itemize}
\item The 2008, 2014, and 2021 UK Research Exercises have been argued to lead to
hiring booms and busts (where initially many new hires are made thereby
temporarily increasing the VCR thereafter increasing the denominator so that
in subsequent years the VCR falls). 
\item The USS pensions scheme went through various changes in response to projected deficits and tax changes that affected retirement decisions and potentially vacancy creation (in 2011 there was an increase in contributions and a career-average scheme for new members; in 2016 there was an increase in contributions and the final salary section was closed to existing members, with a new defined contribution section for higher salaries; in 2018 the proposal to close career-average scheme leads to industrial action). 
\item The big global events like the GFC of 2008, the Brexit referendum of 2016 and its aftermath, the
Trump trade war and deterioration in international relations, and the Covid
spike of 2020 all occurred during the sample period. They all had major
impacts on national and even global macroeconomic quantities so they may
also be expected to have had an effect on university vacancy rates. In particular, following the GFC of 2008 many universities in the US and UK had hiring freezes, which lead to hiring opportunities for those that did not have such freezes or for others as those freezes were relaxed.  
\end{itemize}
The key
point is that all these factors are likely to have differential effects
across universities depending for example on the mix of subjects covered by
different universities, their REF strategies, and their exposure to global
factors. We note that even if a correlation has been established, which is doubtful, such confounding factors make it impossible to claim that evidence has been found that the EJRA is the cause of any effect. For the EJRA to be lawful, objective evidence must be provided by the employer that the policy achieves the aims in a proportionate way and that the aims could not be achieved by other non-discriminatory measures. 

\begin{figure}[]
\centering
\includegraphics[width=0.9\textwidth]{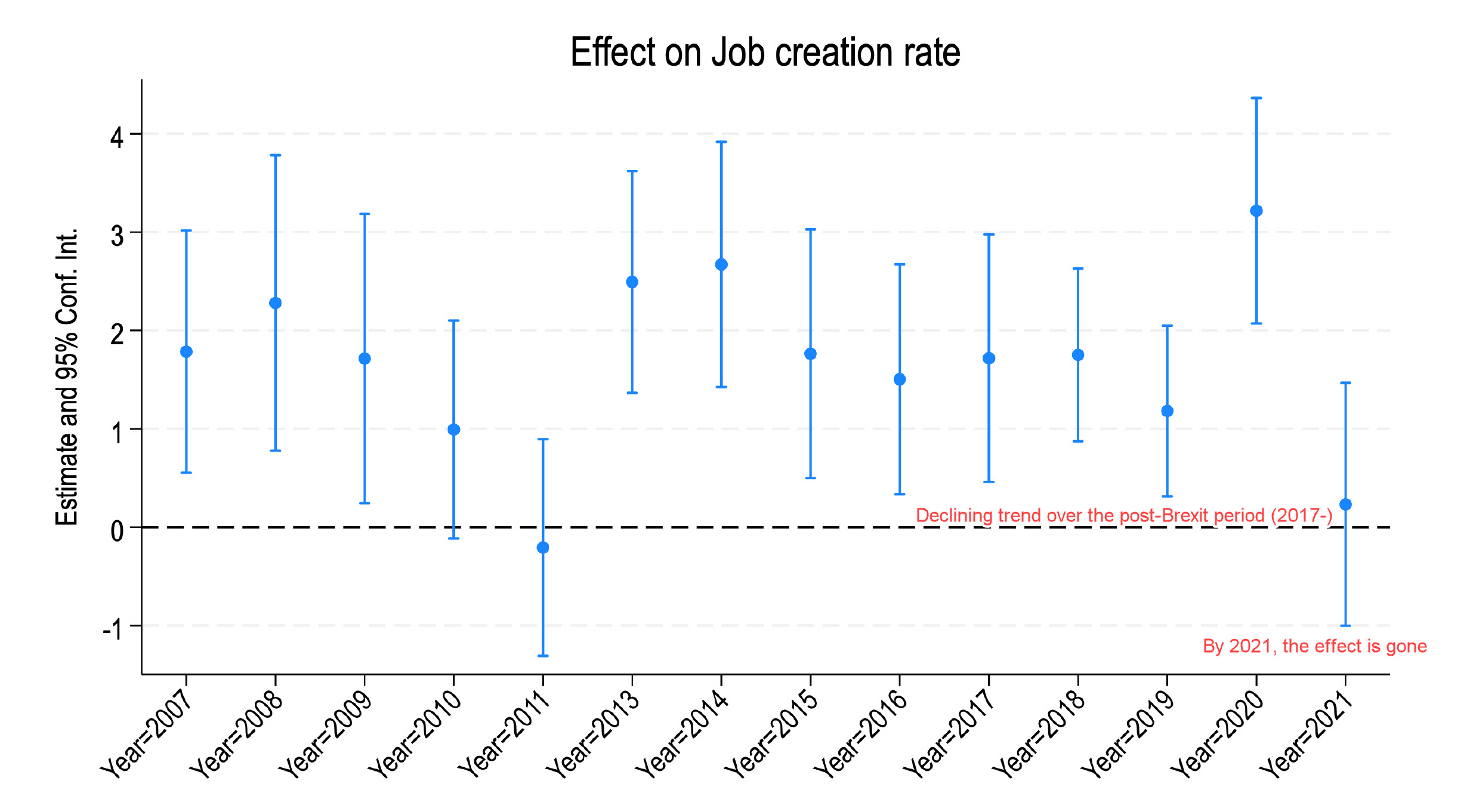}  
\caption{Figure A30 reproduced from HESA Data Report (\protect\cite%
{CambridgeHESA2024})}
\label{fig:PentyFigureA30}
\end{figure}

\subsection{Future Effects of the EJRA Policy}

The Data Report forecasts the future effects of a policy change based on the
dynamical systems model of \textcite{LarsonDiaz2012} that builds on Little's
law of queuing (\textcite{Little1961}) to study the effect of retirement
policy at MIT. This is a more detailed model than that used by Ewart. Little's law essentially says that in a stationary system, the long run average value of
a stock variable (total number of employees) is equal to the product of the
long term average net arrival rate to the system and the average residence
time in the system. Abolishing the EJRA, for example, would lengthen the
steady state time in the system for each employee and would therefore imply,
if the total number of employees is held fixed, that the number of vacancies
and new hires would fall. The system would take time to adjust to the new
long run equilibrium, the so-called steady state, and the so-called dynamical systems approach models this adjustment. \footnote{%
Rather surprisingly, \cite{Holmes24} says: "\textit{Opponents of compulsory retirement prefer simple queuing models such as Little's which can be manipulated by basic assumptions}". This seems to misrepresent their own work, since at the heart of their calculations the steady state is indeed determined by Little's law.}
To build this model, one
has to estimate the exit rates due to lateral moves, death, and retirement
in the absence of a mandate, which is done from the average values observed
in RRG over the 15 years of HESA data.

One implication of this systems model would be that the statistical model in
(\ref{treat}) is misspecified, since it implies that there is a constant
effect on vacancies from the EJRA (consistent with the new steady state) whereas the systems dynamic adjustment model says that the
response to a change takes time to fully play out to the new steady
state. The kind of dynamic responses are shown in their Figure A39, which show some hopping about the steady state line in some cases. Secondly, the systems dynamic model is treated as exact, whereas the key
parameters are uncertain and estimated from data. In that respect, the
estimates reported in the Data Report are subject to wide variability as is
any social science phenomenon you are trying to forecast 40 years hence.
Thirdly, the systems model makes strong predictions about how vacancies and
employment should have responded when mandatory retirement was abolished in
the RRG in 2011, but the actual data do not seem to adjust in the simplistic
way that such an idealized system predicts. No attempt was made to validate this as a predictive model on the experience of the RRG universities after the abolishment of their mandatory retirement.

The model is also flawed in treating everyone in the queue as the same,
failing to account for the differing contributions and income levels
generated by faculty at different career stages. Senior faculty who stay on
the payroll contribute significantly to the Research Excellence Framework
(REF) returns, which in turn bring in pro-rata funding from central sources.
Senior faculty tend to attract more PhD students due to their track records
and projects. Their research funding attracts overheads, which are typically
calculated as a percentage, resulting in more funding for more senior
faculty. The Higher Education Quality Enhancement (HEQE) Quality-Related
(QR) funding is also based on the actual cost of the faculty member, not a
generic FTE. The model errs by taking a ``unit person/Full-Time Equivalent"
approach. The correct approach would consider an FTE at the typical income
level that a faculty member at that career stage generates.

The conclusions drawn by the University's Review Group  are reported in
Table 1 of the Data Report. In the first ten years, there are on average
27.7 fewer new appointments. In the decade from 2053-2062 there are 13.3
fewer vacancies; which is apparently the chosen steady state regime. To put
this in perspective, consider Table 3.1, which shows the total number of
established academic staff at Cambridge over the period 2013-2023. The
fourth column shows that in 2023 a total of 33 members of academic staff
left office at the end of their EJRA year, ostensibly due to the EJRA
itself, rather than to their own personal preferences. However, from 2022 to
2023, the total number of established staff grew from 1642 to 1669, a net
addition of 27. In fact, since 2013, the number of established staff has
grown by approximately 1\% per year. If we extrapolate that trend in
establishment size to 2053/2062, the loss in new vacancies would be more
than matched by the the number of new positions every year. One final point
to bear in mind when looking at the long term steady state, the State
pension age is set to increase to 69 in 2046-2048 due to the increase in
longevity.

\section{Conclusion}

This paper critically evaluates the findings of the Data Report for the EJRA
Review Group, identifying crucial methodological flaws and
misinterpretations. The Data Report suggests that the Employer Justified
Retirement Age (EJRA) at the University of Cambridge increased job creation
rates and contributed positively to the academic workforce. Our analysis
reveals several key issues, including selective filtering applied to the
data, inconsistent treatment of variables, and erroneous conclusions drawn
from statistical models.

The Data Report uses internal HR data from Cambridge for defining groups,
with filters applied to differentiate between EAC Staff and EAR Staff. These
filters include various employment contract details and salary points based
on internal Cambridge HR data, which are not necessarily applicable or
comparable to other UK HEPs. Relevant data on job creation and retirements
are excluded without clear justification, potentially deflating or inflating
the observed job creation rates. Furthermore, the job creation rate is
defined differently for Cambridge and the RRG, resulting in incommensurate
comparisons.

The Data Report also contains inconsistencies in how variables are treated
and measured. The time periods analysed for Cambridge were not directly
comparable to those for the RRG, with trends and changes post-2011 not
uniformly analysed across all institutions. Key variables such as retirement
age and job creation rates are defined and calculated differently, failing
to account for structural differences between institutions. They also fail to account for the difference between vacancies and jobs, and fail to account for the difference between vacancies due to the EJRA and vacancies due to voluntary retirement at that age. Additionally, the Data Report does not uniformly adjust
for macroeconomic factors and other external influences, leading to an
inaccurate representation of EJRA's effect.

Importantly, the Difference-in-Differences model used in the Data Report
does not adequately control for pre-existing trends and differences between
Cambridge and the RRG, resulting in unjustifiable findings on EJRA's impact
on job creation. The Data Report's claim that EJRA led to a 1.6 percentage
point increase in job creation does not hold under proper controls and
comparisons. The dynamical systems model used to forecast the effect of a
policy change is inconsistent with the static regression approach, assuming
a constant effect over time and ignoring potential variable interactions.
The Data Report's results lack of: robustness checks, valid confidence
intervals, and measures of uncertainty, with statistical significance
disappearing when alternative methods (such as the wild bootstrap) were
applied, suggests that their conclusions are unreliable.

The raw data shows that Cambridge University consistently had lower job
creation rates for Established Academic Careers (EAC) compared to the RRG
average, both before and after EJRA implementation in 2011. Cambridge's
average job creation rate was approximately 3\% per year, while the RRG
average was about 5\%. There was no relative improvement in job creation
rates at Cambridge post-EJRA implementation. Every year since 2011,
Cambridge has created fewer EAC jobs than the RRG average, contradicting the
Data Report's claim of a positive impact.

Overall, we can conclude that the EJRA policy did not achieve its intended
goal of increasing job creation for young academics and may have exacerbated
existing disparities in job creation rates compared to other leading
universities. Cambridge consistently had lower job creation rates for
Established Academic Careers (EAC) compared to other Russell Group
universities, both before and after the EJRA implementation, implying that
EJRA is \textit{not} a significant factor driving job creation rates. Other
Russell Group universities, which do not have an EJRA, exhibit higher job
creation rates, suggesting that job creation can be sustained or even thrive
without such a policy.

Cambridge has a poor record on discrimination, going back in history to the
much delayed acceptance of female graduates, and now to continued age
discrimination against academics. As documented in this paper, the Data
Report which underpins EJRA is flawed and its claims about EJRA's
contribution to intergenerational fairness, career progression, and
succession planning are unsupported. The proposal to abolish EJRA for only
academic-related staff is discriminatory against academic staff. Abolishing
EJRA also for academic staff will allow senior academics to continue
contributing substantially to the university's research
income which supports early career researchers. This will foster a fair and
thriving academic community at Cambridge.

\vspace{.5cm}

\textit{We are grateful to Simon Baron-Cohen and Matthew
Kramer for helpful comments. We are also grateful to an anonymous researcher
involved in the HESA data report for answering our questions.}

\vspace{.5cm}

\textit{Dedication. We would like to dedicate this paper to the memory of
Ross Anderson who devoted much of his tireless energy to combatting the
spurious arguments put forth in favour of the EJRA at Cambridge.}

\section{Appendix}

The main model that is used for testing whether a significant effect is present
is given in Section 3.1 of the Data Report, the first displayed equation.\footnote{%
There are several typos in the equations and we have verified by personal communication that this is what they did.}
We can rewrite this equation in more standard notation as 
\begin{equation}
r_{it}^{jc}=\alpha _{i}+\gamma _{t}+\delta D_{it}+\varepsilon _{it}.
\label{treat}
\end{equation}%
The variable $D_{it}$ is the treatment dummy, equal to one for Cambridge
post-2012 and equal to zero for Cambridge pre-2012, and equal to zero for
all the RRG universities from 2007-2008 to 2021-2022. This approach is
called a reversal design and it is used since the retirement policy at
Cambridge did not change, while it did at the RRG (it was abolished). The key feature of this model is that there is no interaction between university and time except through the specific role of $D_{it}$. This is clearly a very strong assumption that they themselves show is not credible. 

The parameter of interest is $\delta $, which captures the effect of EJRA on
job creation (which is implicitly assumed constant across time). The
statistical null hypothesis is that $\delta =0$, i.e., that the EJRA had no
effect on job creation versus the general alternative that it did. Table A.4
column 1 reports the results of this regression separately for EAC and EAR
groups.

The authors compute the so-called difference in differences 
\begin{equation}
\widehat{\delta }=\left( \overline{Cam}_{post}-\overline{Cam}_{pre}\right)
-\left( \overline{\overline{RRG}}_{post}-\overline{\overline{RRG}}
_{pre}\right) ,  \label{a}
\end{equation}%
where the overbar indicates averages over time and the double bar indicates
averages over universities in RRG. It is important to note that there are
only five observations in $\overline{Cam}_{pre}$ and ten observations in $%
\overline{Cam}_{post}$. In Figure A30, they report 
\begin{equation}
\widehat{\delta }_{t}=Cam_{t}-{Cam}_{2012}-(\overline{RRG}_{t}- \overline{RRG%
}_{2012}).
\end{equation}%
In this case there is only one observation in $Cam_{t}$. By comparing with
2012, they are using the point of maximum difference to pivot around. A more
standard approach would be to compare with the pre-policy average $\overline{%
Cam}_{pre}-\overline{\overline{RRG}}_{pre}$.

The authors' main estimates in column 1 of Table A.4. indicate that $\delta$
is positive and statistically significant under their calculated standard
errors. However, when they apply the wild bootstrap standard errors in
Section 3.4, they report that the statistical significance vanishes. The
authors acknowledge that the usual parallel trends assumption is violated in
this data, as is clear from Figures A1 and A30. This would imply the
estimates in the first column of Table A.4 are biased. To address this
issue, they do the following described on page 9. They fit the regression
model 
\begin{equation}
r_{it}^{jc}=a_{i}+b_{i}t+e_{it},
\end{equation}%
for each university using the data before 2012 (5 observations) to obtain $%
\widehat{a}_{i},\widehat{b}_{i}$ and then calculate the residual $\widehat{r}%
_{it}^{jc}=r_{it}^{jc}-\widehat{a}_{i}-\widehat{b}_{i}t$ for all $i,t.$
Then, they take $\widehat{r}_{it}^{jc}$ as the dependent variable in (\ref%
{treat}).

This is absurd. To understand why, imagine constructing a linear trend from
five observations on any noisy time series and then extrapolating a further
ten observations based on that trend. This is extremely bad practice from a
time series point of view, and any public body such as the Bank of England
could not argue policy on the basis of such evidence. In Figure A3, one can
see two RRG universities have rather extreme boosts in job creation followed
by a sharp reversal in one case and slower reversal in the other case, all
before 2012. So whatever led them to increase job creation and decrease job
creation could be repeated at any time after 2012. Logically, if you
extrapolated the trend backwards, it appears that for some universities in
2000, the job creation would be negative. Hence the foundation of this
pretrend assumption/method is very weak. Mechanically, the result of this method is to drag down the RRG values
as those fake pretrends did not continue and this gives a boost to the
estimated treatment effect. The treatment effect here is to be interpreted
relative to a pre-established trend, so the number 4.408 in Table A.4 column
2 is not directly comparable with the result from column 1. In addition, the
standard errors are calculated incorrectly since they ignore the first stage
detrending as if this has no effect on the sampling variability - but it has
a huge effect on standard errors so these results are misleading and
deceptive.

The second method they adopt to address the pre 2012 \textquotedblleft
trend" is based on the regression 
\begin{equation}
r_{it}^{jc}=\alpha +\beta s_{it}+u_{it},
\end{equation}%
where $s_{it}$ is the log of FTE student numbers at university $i$ in year $t
$. In this case, the $\beta $ is assumed constant over universities and
again the estimation of $\beta $ is done using the pre-2011 data. They then
calculate $\widetilde{r}_{it}^{jc}=r_{it}^{jc}-\widehat{\beta }s_{it}$ for
all $i,t$ and then apply the DID to this residual. The same comments apply
to this method in that the preliminary regression on student numbers for
five years of data is likely to be very noisy and this noise is not properly
taken account of in the standard errors. 

Also, the specification seems a bit dubious to say the least. Why use only pre EJRA data to estimate the effect of student numbers on vacancies? Why restrict the parameter $\beta $ to be constant across all universities when we know that student staff ratios vary considerably across universities. In fact, we have done some detective work ourselves to address \cite{Holmes24}. We managed to reconstruct the VCR time series for Cambridge and the RRG from their Figure A1 using specialized software and obtained the student numbers by university and academic year directly from HESA. In our reconstruction of their regression we find that the pooled slope is positive, whereas the slope for Cambridge separately is negative and the slope for RRG separately is positive (none is statistically significant though even if you take the OLS standard errors), which is what they work with, apparently. This is why relative to that pre EJRA trend, Cambridge appears to do well compared to RRG. 
It is also odd to
use the log of student numbers to predict the ratio of new hires. If we
assume that the staff-student ratio is targeted to be constant we would have
(where $N$ is the number faculty and $S$ is student numbers): 
\begin{equation}
0=d\left( \frac{N}{S}\right) =\frac{dN}{S}-\frac{N}{S}\frac{dS}{S}=\frac{N}{S%
}\left( \frac{dN}{N}-\frac{dS}{S}\right) ,
\end{equation}%
so that $dN/N$ should be equal to $d\log S$, not $\beta \log S.$
Empirically, $s=\log S$ is more related to $\log N$ than $d\log N$. The estimates from this equation are therefore not credible either.
Interestingly, in column 3 of Table A.4 they show that for this method the
EJRA for EAR staff produces a significant \textit{increase} in vacancies. If
these are the results they wish to hang their clogs on then they would have
to change their proposal to abolish EJRA for EAR staff. 

For inference, they are using normal critical values in the main part, which can be justified if the difference of vacancy ratios is exactly normally distributed or the relevant sample size is large; here, the binding case is the pre-EJRA period, which has only five observations, which is not usually considered enough for a Central Limit Theorem. The inference methods are \textquotedblleft robust standard errors" in the
main part (although what they are robust to is not spelled out) and a wild
bootstrap method in Section 3.4. The standard errors in Table A.4
effectively assume that $\varepsilon _{it}$ is independent across $i$ and $t,
$ that is, after accounting for the \textquotedblleft main effects" $\alpha
_{i}$ (university specific effect) and $\gamma _{t}$ (time specific effect),
the remaining movements are uncorrelated. 
This is clearly a bad assumption
in this case as the universities in the RRG are not separate individuals
whose health outcomes do not affect others' outcomes but directly competing
institutions who are subject to the same rules and procedures (apart from
EJRA). The pre-trend estimation is not taken into account of in the standard
errors. Instead, the authors just supply the residuals to STATA and get the
usual standard errors that would arise if there were no pre-estimation.
Clearly, estimating two parameters from five observations is pushing the
boundaries of credibility. Credible standard errors would be much larger
than those presented. Indeed, as acknowledged in Section 3.4 (robustness
checks), when the wild bootstrap is used (this takes account of
heteroscedasticity but not cross university and cross time correlations),
the statistical significance disappears. Likewise, when a synthetic control
method is used, that is, finding a linear combination of RRG that best
matches Cambridge in the pre treatment period with respect to vacancy rate,
they report that the statistical significance disappears.

\printbibliography
\end{document}